\documentclass[manuscript]{aastex}
\usepackage{emulateapj5}

\newcommand{\umu}{\mu}

\newcommand{\myemail}{chris.willott@nrc.ca}

\newcommand{\Msolar}{\mbox{\,$\rm M_{\odot}$}}        

\newcommand{\mg}{Mg\,{\sc ii}}
\newcommand{\fe}{Fe\,{\sc ii}}
\newcommand{\civ}{C\,{\sc iv}}
\newcommand{\lya}{Ly\,$\alpha$}
\newcommand{\nv}{N\,{\sc v}}
\newcommand{\hbeta}{H\,$\beta$}

\shorttitle{A $3 \times 10^{9}$  solar mass black hole in  SDSS J1148+5251}
\shortauthors{Willott et al.}

\begin{document}

\title{A $3 \times 10^{9}$ solar mass black hole in the quasar SDSS J1148+5251 at $z = 6.41$}

\author{Chris J.\ Willott} 
\affil{Herzberg Institute of Astrophysics, National Research Council,
5071 West Saanich Rd,\\ Victoria, B.C. V9E 2E7, Canada}
\email\myemail

\author{Ross J. McLure}
\affil{Institute for Astronomy, University of Edinburgh, Royal
Observatory,  Blackford Hill, Edinburgh, EH9 3HJ, U.K.}
\email{rjm@roe.ac.uk}

\and

\author{Matt J. Jarvis}
\affil{Astrophysics, Department of Physics, Keble Road, Oxford, OX1
3RH, U.K.}
\email{mjj@astro.ox.ac.uk}

\begin{abstract}

We present near-infrared $H$ and $K$-band spectra of the $z=6.41$
quasar SDSS J114816.64+525150.3. The spectrum reveals a broad
\mg\,$\lambda 2799$ emission line with a full-width half-maximium of
6000 km\,s$^{-1}$. From the peak wavelength of this emission line we
obtain a more accurate redshift than is possible from the published
optical spectrum and determine a redshift of $z=6.41 \pm 0.01$. If the
true peak of the \lya\, emission is at the same redshift, then a large
fraction of the flux blueward of the peak is absorbed. The equivalent
width of the \mg\, emission line is similar to that of lower redshift
quasars, suggesting that the UV continuum is not dominated by a beamed
component. Making basic assumptions about the line-emitting gas we
derive an estimate for the central black hole in this quasar of $3
\times 10^{9}$ solar masses. The very high luminosity of the quasar
shows that it is accreting at the maximal allowable rate for a black
hole of this mass adopting the Eddington limit criterion.
\end{abstract}


\section{Introduction}

The vast energy requirements of the most luminous quasars can be met
by a model invoking extraction of gravitational potential energy from
matter falling toward a supermassive ($\sim$ a billion solar mass)
black hole. The correlation between black-hole mass and galaxy mass
observed in the local Universe strongly suggests that the most
luminous quasars will reside in the most massive galaxies in the most
massive dark matter halos. The luminous quasars at redshifts $z>6$
found in the Sloan Digital Sky Survey (SDSS) therefore pinpoint the
earliest massive objects to form (Fan et al. 2001, 2003).

The recent discovery of SDSS J114816.64+525150.3 (hereafter SDSS
J1148+5251) at a redshift of $z=6.43$ by Fan et al. (2003) makes it
the most distant known quasar, observed only 840 million years after
the beginning of the universe (cosmological parameters of $H_0=70~ {\rm
km~s^{-1}~Mpc^{-1}}$, $\Omega_{\mathrm M}=0.3$ and $\Omega_\Lambda=0.7$
are assumed throughout). The very high luminosity of this quasar
($M_{1450} =-27.8$) suggests that it contains a black-hole mass of at
least several billion solar masses.

We have obtained near-infrared spectroscopy of SDSS J1148+5251 to
search for the \mg\, emission line (rest wavelength 2799\,\AA)
redshifted into the $K$-band. This line is particularly useful since
it is usually found close to the systemic redshift, unlike higher
ionization lines such as \civ\, which typically show large blueshifts
$\sim 1000$ km\,s$^{-1}$ (Richards et al. 2002). The redshift for this
quasar given by Fan et al. was based on the position of a heavily
absorbed \lya\, line and the onset of the Lyman break and is
therefore quite uncertain. An accurate redshift is important for
determining the strength of the proximity effect and for molecular
line searches.  Additionally, it has recently been shown that the
width of the \mg\, emission line can be used to derive the mass of the
quasars central black hole in the same way that the \hbeta\, line has
traditionally been used (McLure \& Jarvis 2002).

We describe the spectroscopic observations and measurements made from
them in \S 2. The importance of the new redshift determination is
discussed in \S 3.  We determine an estimate for the mass of the black
hole in \S 4.  Our conclusions are presented in \S 5.

\begin{figure*}
\plotone{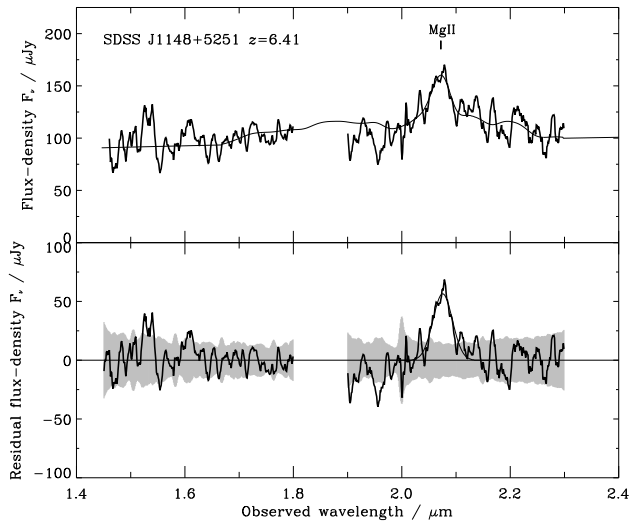}
\vspace{0.9cm}
\caption{{\it Upper:} UIST $H$ and $K$-band spectrum of SDSS
J1148+5251 (thick line; smoothed with a 7 pixel boxcar). The location
of the broad \mg\, emission line is marked. The smooth curve shows the
best fit comprising a power law continuum ($\alpha=0.2$), a broadened
\fe\, template and a broad \mg\, line.  {\it Lower:} The residual
spectrum after subtraction of the power law continuum and \fe\,
template (thick line). The best fit to the broad \mg\, doublet
(FWHM=6000 km\,s$^{-1}$) is also shown here (thin line). The shaded
region shows the $\pm 1\sigma$ noise level of the observation.
\label{fig:spec}}
\end{figure*}

\section{Near-infrared spectroscopy}

\subsection{Observations}

We observed SDSS J1148+5251 with the United Kingdom Infra-Red
Telescope (UKIRT) equipped with the UKIRT Imager Spectrometer (UIST)
on 2003 January 10. UIST is a 1-5 $\umu$m imager-spectrometer with a
1024x1024 InSb array. Conditions were photometric with seeing at
$K$-band approximately 0.6 arcsec. The $HK$ grism was employed with a
0.48 arcsec wide slit giving continuous wavelength coverage from 1.4
to 2.5 microns with a resolving power of $\sim 500$. The pixels have
size 10.9 \AA\ spectrally and 0.12 arcsec spatially. The target was
nodded along the slit a distance of 12 arcsec in between exposures to
enable accurate sky subtraction. The total integration time of the
observation was 6480 seconds.

The data frames were flat-fielded and then adjacent pairs were
subtracted from each other to subtract the sky emission and produce
positive and negative spectra of the target. These data were combined
and any pixels deviating from the mean at the $ 3 \sigma$ level were
excluded since this indicates contamination by cosmic rays. The
residual background level was subtracted from the combined image. The
positive and negative quasar spectra were extracted from this image
using apertures of 1.0 arcsec and combined. 

Flux calibration and atmospheric extinction corrections were made by
dividing the spectrum by that of a star of spectral type F6 which was
observed immediately preceding the quasar. An accurate wavelength
calibration was made by comparison to an Argon arc lamp. This solution
was then checked against the positions of about 20 sky emission lines
giving an rms uncertainty in the wavelength calibration of 1\,\AA. The
absolute flux scale of the spectrum was set by comparison with the
magnitude measured in a snapshot $K$-band image. The image gave
$K=16.86 \pm 0.05$ consistent with the $K'=16.91$ quoted by Fan et
al. (2003).

\subsection{Emission line fitting}

The spectrum of SDSS J1148+5251 is shown in Fig.\ \ref{fig:spec}. It
shows a broad emission line at approximately the position expected for
redshifted \mg. To determine the wavelength and width of the \mg\,
emission line, a line-fitting process was applied which involves the
determination of the underlying continuum and \fe\, emission
surrounding the \mg\, line. Due to the blending of the \mg\, line
emission and the \fe\, features in the 2700\,\AA\ -- 2900\,\AA\
region, the fitting of the continuum and \fe\, emission is confined to
two bands either side of the \mg\, line (rest-frame 2250\,\AA\ --
2700\,\AA\ and 2900\,\AA\ -- 3100\,\AA). 

The \fe\, emission in this spectral region is modeled using an \fe\,
template based on an archival HST FOS spectrum of the narrow line
Seyfert galaxy I\,Zw\,1, which is notable for its extremely strong
\fe\, emission. Due to the fact that the \fe\, emission in I\,Zw\,1 is
relatively narrow (FWHM $\sim 900$ kms$^{-1}$), this template can be
smoothed in order to match the much broader \fe\, features typical of
powerful quasars. The fitting of the \fe\, emission template and AGN
continuum are performed simultaneously, with the amplitude and FWHM of
the \fe\, iron template left as free parameters. The AGN continuum is
modeled as a power law which is normalized with the continuum
magnitude at rest-frame 1280\,\AA\ of $AB_{1280}=19.10$ given by Fan
et al. (2003). The spectral index of the power law is a free parameter
and the fitting produces a value of $\alpha=0.2$ (defining $f_{\nu}
\propto \nu^{-\alpha}$, where $f_{\nu}$ is the flux-density at
frequency $\nu$). This continuum slope is typical of quasars
(e.g. Vanden Berk et al. 2001).  The goodness of fit for the
\fe\,+continuum fit is determined from a chi-square test using the
noise array from the observation.

Once the minimum chi-square fit has been determined the best-fitting
combination of continuum and \fe\, emission is then subtracted from
the spectrum, leaving the isolated \mg\, emission line in the region
2700\,\AA\ -- 2900\,\AA. The \mg\, emission line itself is then
modeled using two identical gaussians which represent the line doublet
($\lambda \lambda 2796, 2802$). During the fitting of the line
profile, the amplitude, FWHM and central wavelength of the gaussians
are treated as free parameters, while the doublet separation is held
fixed at its laboratory value.  The best fit to the \mg\, emission was
determined with a chi-square test. Although more complex
deconvolutions of the line profile are obviously possible, due to the
signal-to-noise constraints of the spectrum the above procedure was
adopted because it was able to effectively describe the \mg\, line
profile with the minimum level of complexity. 

We also performed an alternative fitting procedure in which the
continuum was fixed at two points with low expected \fe\, emission and
then the \mg\, and \fe\, features fit simultaneously. The results from
this independent fitting method are consistent with those using the
former method. The ratio of \fe\,/\mg\, emission lines in high
redshift quasars is a powerful diagnostic of early element production
inside supernovae (Yoshii, Tsujimoto \& Kawara 1998). Unfortunately we
do not have sufficient signal to noise to place meaningful constraints
on this ratio using either of our two fitting methods.

The best fit \mg\, line doublet in SDSS J1148+5251 is at a redshift of
$z=6.41 \pm 0.01$. The deconvolved FWHM of each component of the
doublet is $6000\, ^{+1100}_{-600}$ km\,s$^{-1}$. The rest-frame
equivalent width of the \mg\, line is 35\,\AA. The typical equivalent
width of \mg\, lines in lower redshift SDSS quasars is 32\,\AA\
(Vanden Berk et al. 2001). The fact that the equivalent width of \mg\,
in SDSS J1148+5251 is similar to that of other quasars indicates that
the UV continuum emission from this quasar is not strongly beamed. Fan
et al. (2003) noted that the \lya\, equivalent width of this and other
$z\sim6$ quasars are smaller than lower redshift quasars. Pentericci
et al. (2002) observed the \civ\, line in the $z=6.28$ quasar SDSS
J1030+0524 and also found an equivalent width typical of lower
redshift quasars. These results suggest that the low \lya\, equivalent
widths are primarily due to \lya\, absorption and not a fundamental
difference between the emission properties of these quasars.

\section{An accurate redshift determination}

The redshift of $z=6.43 \pm 0.05$ for this quasar given in Fan et
al. (2003) is based on the position of the heavily absorbed \lya\,
line and the onset of the Lyman break. The lack of distinguishable
\nv\, emission or any other lines in the optical spectrum mean that a
more accurate redshift could not be obtained. In our spectrum we are
able to measure the location of the \mg\, line and hence obtain a more
accurate redshift for this quasar. We find a redshift of $z=6.41 \pm
0.01$.  The \mg\, line is particularly useful for redshift estimation
compared to, for example, \civ\, which could be observed in the
$J$-band. The low-ionization \mg\, line is usually found close to the
systemic redshift, whereas the high-ionization \civ\, line often shows
blueshifts of several thousand km\,s$^{-1}$ (Richards et al.
2002). Another important fact is that there is no systematic velocity
offset for the \lya\, and \mg\, lines in a large sample of SDSS
quasars (Vanden Berk et al. 2001), so we can predict the intrinsic,
unabsorbed \lya\, line peak.

Accurate redshifts for the most distant quasars are important for
several reasons. The rest-frame UV spectra of $z\sim 6$ quasars show a
deficit of flux shortward of \lya\, due to absorption by neutral
hydrogen in the intergalactic medium (Fan et al. 2001; Becker et
al. 2001). These observations have led to the possibility that we are
now witnessing the epoch of reionization. Whilst accurate quasar
redshifts are not essential for determining the ionization state of
the universe as a function of redshift, they are very important when
dealing with the state of the IGM close to the quasar. 

Assuming that the IGM in general is neutral at the redshift of the
quasar, the strong ionizing field of the quasar will ionize a local
region around the quasar allowing \lya\, photons to be
transmitted. This {\it proximity effect} has been observed as a lack of
absorption lines in the spectra of quasars close to the peak of
\lya\, (Bajtlik, Duncan \& Ostriker 1988). Cen \& Haiman (2000)
showed that the size of this ionized region and the transmitted flux
blueward of the peak of \lya\, depend upon the number of ionizing
photons which have been emitted from the quasar, i.e. the product of
its lifetime and luminosity. Haiman \& Cen (2002) apply this to the
spectrum of the $z=6.28$ quasar SDSS J1030+0524 to show that it is not
strongly gravitationally lensed and to derive a lower limit on how
long the quasar has been active. Such an analysis depends critically
on knowing where the peak of the \lya\, line should be in the
spectrum.

The spectrum of SDSS J1148+5251 presented by Fan et al. (2003) also
shows zero flux at \lya\, redshifts of 5.7 to 6.33, suggestive of a
neutral universe. However, there is some flux apparent on the blue
wing of \lya\, which is likely due to the proximity effect. The marked
location of \lya\, at $z=6.43$ given by Fan et al. is at the peak of
the \lya\, emission. However, adopting the new redshift for the quasar
of $z=6.41$ shifts the location of \lya\, to 9010\,\AA, which is
co-incident with a steep decline in flux to less than half of the peak
value. Hence the revised redshift indicates a substantially more
neutral IGM close to the quasar. The amount and extent of transmitted
flux on the blue side of \lya\, is similar to that in SDSS J1030+0524
and the conclusions about this quasar made by Haiman \& Cen (2002) and
Pentericci et al. (2002) probably apply equally to SDSS J1148+5251.

The evolutionary state of the host galaxies of high redshift quasars
can be probed by (sub)millimeter observations of dust and molecular
gas. Particularly important tracers are the emission lines from
transitions of CO molecules and carbon atoms (Blain et al. 2000).
However current millimeter receivers have a fairly small bandwidth
corresponding to $\sim 2000$ km\,s$^{-1}$. Hence systemic redshifts
must be known to an accuracy of $\sim 0.01$ to ensure that the
emission line will fall within the frequency range of the
observation. Our observations of SDSS J1148+5251 provide a redshift of
sufficient accuracy for millimeter line searches.

\section{Estimating the black-hole mass}

Reverberation mapping studies of low redshift quasars have been
successfully employed to estimate the masses of their black holes
(Netzer \& Peterson 1997). This method determines the radial distance
of the broad line emitting region (BLR) from the nucleus ($R_{\rm
BLR}$), by observing the time lag ($\sim 1$ year) between variations
in the UV/optical continuum and the line strength. Combining this
distance with the velocity of the emission line gas ($V_{\rm BLR}$,
which is closely related to the observed FWHM of the line) gives a
virial estimate of the black-hole mass ($M_{\rm bh}= G^{-1}R_{\rm BLR}
V_{\rm BLR}^2 $). This estimate assumes that the dynamics of the gas
are dominated by gravitational forces (see Peterson \& Wandel 2000 for
a discussion). These reverberation studies have revealed a correlation
between the continuum luminosity and the radius of the BLR which can
be used to estimate black-hole masses when only the line FWHM and
luminosity are available (Kaspi et al. 2000).

Traditionally, the \hbeta\, emission line has been used for this work,
since it is readily observable in the optical spectra of low redshift
quasars where the reverberation mapping calibration of luminosity and
BLR has been derived. Recently, McLure \& Jarvis (2002) have
cross-calibrated this relationship from the \hbeta\, line to the \mg\,
line using a sample of AGN with reverberation mapping measures of
$R_{\rm BLR}$, so that black-hole mass estimates can be derived to
higher redshifts. They note that the \mg\, line (which has an
ionization potential close to that of \hbeta) is a much better choice
than higher ionization lines like \civ\, whose dynamics may be
strongly influenced by outflows. McLure \& Jarvis derived a
calibration between $R_{\rm BLR}$ and $L_{3000}$ (the monochromatic
luminosity at 3000\,\AA) and performed a fit to the \hbeta\, and \mg\,
FWHMs, finding a linear relationship with virtually identical
normalization.  The resulting black-hole mass estimator has the form

\begin{equation}
\frac{ M_{\rm bh}} {\Msolar}  =3.37\left(\frac{\lambda
L_{3000}}{10^{37}{\rm W}}\right)^{0.47}\left(\frac{\rm FWHM(Mg\,II)}
{{\rm kms}^{-1}}\right)^{2}.
\label{final}
\end{equation}

Our spectrum covers the rest wavelength region 3000\,\AA\ and we have
determined the \mg\, FWHM. Therefore we can apply this formula to derive
an estimate of the black-hole mass in SDSS J1148+5251. Using $\lambda
L_{3000}=6.2 \times 10^{39}$ W and FWHM(\mg) = 6000 km\,s$^{-1}$ we get
$M_{\rm bh}=3 \times 10^9$ \Msolar. Assuming that $z\sim6$ quasars
are not fundamentally different to low redshift quasars (evidence
supporting this is that their spectra are similar), the accuracy of
this black-hole mass estimator with respect to those from
reverberation mapping is a factor of 2.5 ($1\sigma$). 

This is an extremely massive black hole to be in existence such a
short time (840 Myr) after the beginning of the universe. Assuming
that this black hole is residing in a proportionately massive halo
($\sim 10^{13}$ \Msolar) then standard theories of structure formation
predict that this is one of the rarest, most massive halos to have
collapsed by this time (e.g. Haiman \& Loeb 2001; Fan et al. 2001).
There are two caveats to the derived black-hole mass. Since the
estimator depends on $L_{3000}^{0.47}$, if we have over-estimated the
intrinsic, isotropic luminosity we will have over-estimated the black
hole mass. If the luminosity function of $z\sim 6$ quasars has a steep
slope then there are expectations that a significant fraction of the
SDSS $z\sim 6$ quasars will have their luminosities enhanced by either
gravitational lensing or beaming (Turner 1991; Wyithe \& Loeb
2002a). Fan et al. (2003) discussed $K$'-band images of the three new
$z\sim 6$ quasars and found that none of them show evidence for
gravitational lensing. Our spectrum found that the equivalent width of
the \mg\, line is typical of quasars which shows that the continuum
luminosity is not strongly enhanced by beaming. Therefore, the
available evidence suggests that there is minimal amplification of the
intrinsic quasar luminosity.

It is interesting to consider the maximum luminosity that can be
attained by an accreting black hole of this mass. This is usually
considered to be the Eddington limit at which the outward radiation
pressure equals the inward gravitational attraction. The Eddington
limit for a black hole of mass $3\times 10^{9}$ \Msolar \, is $4
\times 10^{40}$ W. Using a bolometric correction of 7 to get the
bolometric luminosity from $\lambda L_{3000}$ (Wandel, Peterson \&
Malkan 1999) we determine a bolometric luminosity for SDSS J1148+5251
of $L=4 \times 10^{40}$ W. Therefore we find that the black hole in
this quasar is radiating the maximal amount possible. This supports
the often made assumption when relating quasar luminosities to black
hole and halo masses that the highest redshift quasars are accreting
at the Eddington limit (e.g. Fan et al. 2001; Wyithe \& Loeb
2002b). It is not too surprising to find that SDSS J1148+5251 has the
minimum size black hole given its luminosity. As mentioned above, the
short cosmic time available for the growth of supermassive black holes
by this redshift means black holes with such masses will only have
arisen within rare collapsed peaks. In such conditions, there is
likely to be a plentiful gas supply to fuel the quasar and the black
hole is expected to be accreting close to its Eddington limit.

\section{Conclusions}

We have presented $H$ and $K$ spectra of the quasar SDSS J1148+5251
which currently has the highest known redshift. We clearly detect the
broad \mg\, emission line. Fitting the location of this emission line
we derive a redshift of $z=6.41 \pm 0.01$. The equivalent width of
this line is similar to that of lower redshift SDSS quasars,
suggesting that the optical continuum is not strongly beamed. We
obtain a virial estimate of the black-hole mass of $M_{\rm bh}=3
\times 10^9$ \Msolar.  The quasar is radiating at the Eddington
luminosity, in line with the expectation that it is easier to fuel a
supermassive black hole than create one at such an early
epoch. Similar observations of other $z>6$ quasars would reveal if
they are all radiating at the Eddington luminosity.

\acknowledgments

We thank the staff at the UKIRT for their excellent technical
support. The United Kingdom Infrared Telescope is operated by the
Joint Astronomy Centre on behalf of the U.K. Particle Physics and
Astronomy Research Council. We thank the referee for comments. CJW
thanks the National Research Council of Canada for support.

\end{document}